# A CASE STUDY COMPARING THE COMPLETENESS AND EXPRESSIVENESS OF TWO INDUSTRY RECOGNIZED ONTOLOGIES

**Highlights**

- Comparison of Brick and Project Haystack ontology for use with Smart Building applications
- Qualitative document review of ontologies
- Quantitative expressiveness and completeness assessment of ontologies

# A CASE STUDY COMPARING THE COMPLETENESS AND EXPRESSIVENESS OF TWO INDUSTRY RECOGNIZED ONTOLOGIES

Caroline Quinn[1], J.J. McArthur[1]*

[1] Department of Architectural Science, Ryerson University

350 Victoria St., Toronto, ON, M5B 2K3

*corresponding author: jjmcarthur@ryerson.ca

(phone : +1 416-979-5000 x554082 fax: +1-416-979-5353 )

**Abstract**

Enabling Smart Building applications will help to achieve the ongoing efficient commissioning of buildings, ultimately attaining peak performance in energy use and improved occupant health and comfort, at minimum cost. For these technologies to be scalable ontology must be adopted to semantically represent data generated by building mechanical systems, acting as conduit for connection to Smart Building applications. As the Building Automation System (BAS) industry considers Brick and Project Haystack ontologies for such applications, this paper provides a quantitative comparison of their completeness and expressiveness using a case study. This is contextualized within the broader set of ontological approaches developed for Smart Buildings, and critically evaluated using key ontology qualities outlined in literature. Brick achieved higher assessment values in completeness and expressiveness achieving 59% and 100% respectively, as compared to Haystacks 43% and 96%. Additionally, Brick exhibited five of six desirable qualities, where Haystack exhibited only three. Overall, this critical analysis has found Brick to be preferable to Haystack but still lacking in completeness; to overcome this, it should be integrated with other existing ontologies to serve as a holistic ontology for the longer- term development of Smart Building applications. These will support innovative approaches to sustainability in building operations across scales and as next- generation building controls and automation strategies.

## 1 Introduction

Smart Building applications are receiving an increasing amount of attention within the Architecture Engineering and Construction (AEC) industry. These applications use data generated by building data sources, allowing for scalable oversight required for the implementation of Smart Building applications. Purpose-driven Smart Building applications such as fault detection and energy optimization of controls, or information access driven tools such as parametric design and Building Information Models for Facilities Management (BIM for FM) are all examples of Smart Building applications. Smart Building applications are of value to a host of building stakeholders including FM, building owners, operators, and tenants. These applications allow for testing and planning of maintenance projects, observation of building conditions, retrieval of specific sensor data, optimize controls strategies, ultimately reducing energy use and improving occupant comfort. Smart Building applications are being developed by Building Automation Systems (BAS) vendors who are adapting their systems to include domain semantics. Despite the plethora of potential ontological approaches developed to represent BAS semantic concepts, there lacks an accepted standardized means to do so. This presents a significant challenge for research and development of Smart Building applications.

While a variety of ontologies have been proposed to define AEC data using a variety of paradigms, this research specifically compares two ontologies of interest to BAS developers: Project Haystack ("Haystack") and Brick which use different paradigms. These ontologies are widely considered to be particularly relevant to Smart Building applications [1, 2, 3]. There is a paucity of research in the comparison of these ontologies, a pertinent area of research as there exists a “tradeoff between expressiveness (what can be said) and inference (what can be obtained from the information represented) in traditional [Haystack] and web-based [Brick] ontology languages.” [4]. By evaluating these ontologies using an industry dataset used by multiple vendors for testing BAS applications, this paper contributes to the ongoing discourse of these ontologies, updating and expanding on previous studies [5] of their earlier versions, and placing them within the broader context of ontological approaches. Ultimately this research and further comparison of ontologies will identify the most promising standardized approach, supporting shared conceptualizations through a common ontology, and allowing the development of Smart Building applications to move forward more efficiently. Such a standardized approach will allow applications to be supplied by a variety of vendors, as well as be updated to meet the needs of system autonomy, maintenance tracking, human system interaction, reporting, *etc.*. The investigation evaluates two ontologies (Brick and Haystack), one of which supports World Wide Webs Consortiums (W3C) semantic web standards, for their suitability to semantically represent BAS data through quantitative completeness and expressiveness measures as well as a qualitative review.

## 2 Literature Review

Semantic concepts are the interpretations of domain concepts within a context such as BAS systems [6]. Ontologies have been specifically conceptualized to represent the semantic concepts of a specific domain. The term ‘ontology’ can be defined as “vocabularies of representational terms - classes, relations, functions, object constants with agreed-upon definitions, in the form of human readable text and machine-enforceable, declarative constraints on their well-formed use.” [7]. An ontology must be specified in a well-defined language [8] – an *ontology specification* *language* – to create shared conceptualizations. Fundamentally different knowledge representation paradigms have been used to form ontology specification languages [4], notably the description logic and object-centered approaches. The object-centered paradigm approach "specify concepts by using a set of restrictions on them" using classifiers [4] , and focus on class membership [9]. Object-centered paradigm approaches use name value pairs with unique identifiers, and focus on identifying instances of concepts [9]. A fundamental difference between the paradigms is that object-centered can allow "two or more descriptions to refer to the same entity" [9] , where a description logic will only allow one description per entity due to tight classifications.  A significant number of ontologies using these paradigms in the BAS domain exist that could be applied to Smart Building applications, and these are presented in Figure 1.

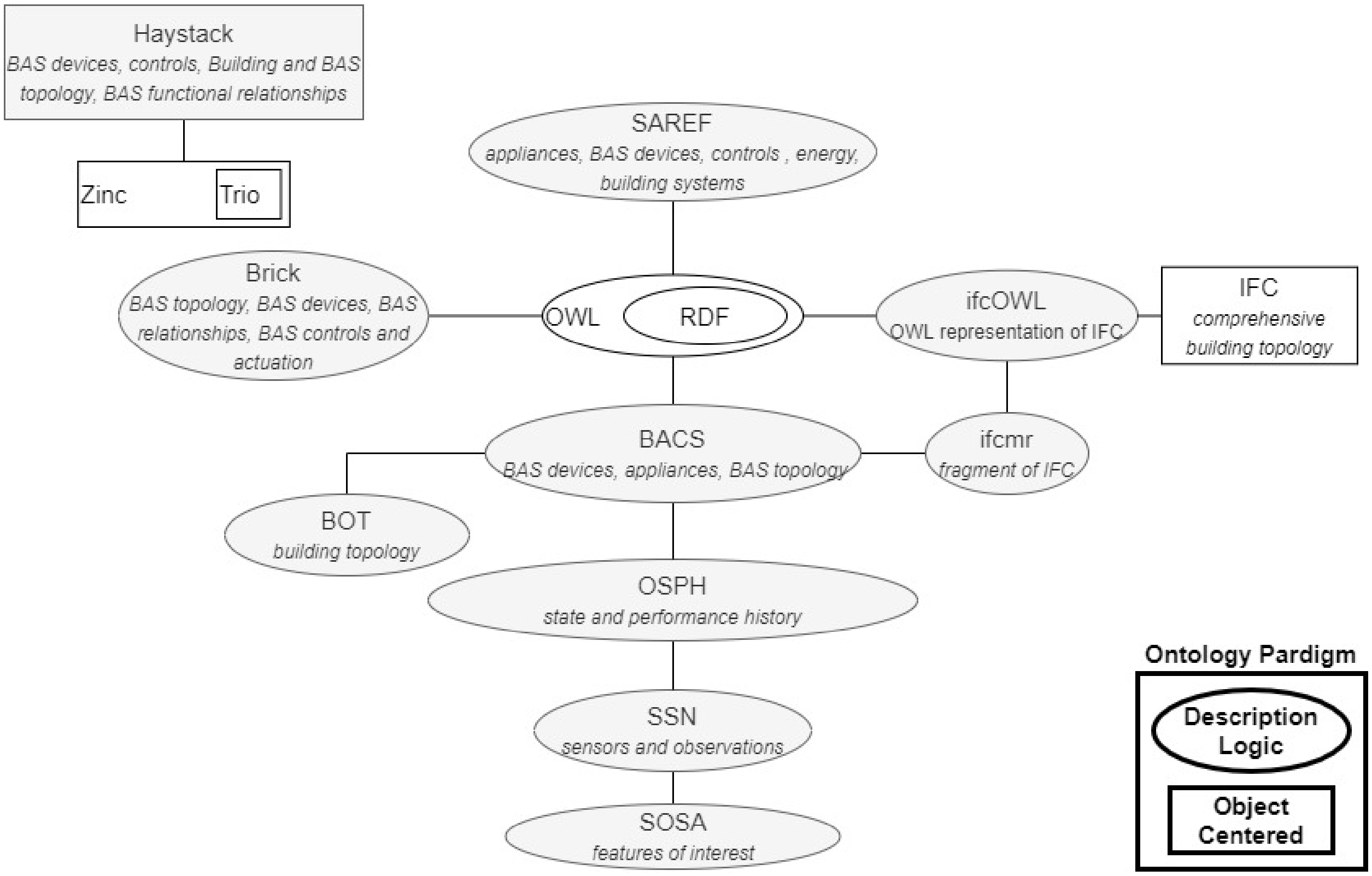

Figure 1 :  BAS ontologies

An object-centered approach can be used for ontology and is classified as either frame-based or object oriented [10]. The Industry Foundation Class (IFC) is an object-centered ISO standard taken by the buildingSmart community, IFC, consisting of a file format and schema used to represent AEC data [11]. IFC was created for information exchange within the AEC industry and has been extensively developed. However, because the IFC schema cannot represent formal semantics, be extended by individual users, or integrate with other web data a variation known as ifcOWL [11] has been represented by the Web Ontology Language (OWL). OWL is a Semantic Web language grounded in description logic and specified by W3C [12]. The other significant ontology using an object-centered paradigm is Haystack. This ontology uses a tagging model and ontology specification language called Trio that is object-centered and developed for Smart Building applications [13] and has become popular in industry [14], and is described in detail in section 2.1. IFC has been compared to Haystack in the past and it was found to have less BAS systems vocabulary coverage but meet the needs of more Smart Building applications [5].

Alternative to object-centered, description logic ontology languages use logic-based knowledge representation that “describe knowledge in terms of concepts and role restrictions that are used to automatically derive classification taxonomies” [10]. The W3C specified the Resource Description Framework (RDF), which “provides two important contributions [to ontologies]: a standardized syntax for writing ontologies and a standard set of modeling primitives, like instance of and subclass of relationships.” [10]. Within the buildingSmart community, the Linked Building Data (LBD) group aimed to create a central ontology using RDF, the Building Topology Ontology (BOT) [11]. BOT operates within the W3C standards and integrate with other industry specific ontologies acting as a typology and linking AEC related ontologies. Some notable BOT adjacent ontologies relevant to the research of BAS ontologies include Semantic Sensor Network (SSN) used to represent sensors actuators and their observations, and the Sensor, Observation, Sample, and Actuator (SOSA) ontology developed to improve the SSN ontology and appeal to a wider audience of users [15]. The LBD group created the Building Automation and Control System ontology (BACS) specifically focused on representing BAS semantics reusing these adjacent ontologies [16]. All BOT ontologies use the W3C description logic RDF and OWL standards. Outside of BOT, the Smart Appliance REFerence Ontology (SAREF) is used to represent sensor data and household appliances also uses OWL and could therefore be manually integrated with BOT if desired. The Brick ontology is another that uses this paradigm and has been considered by industry for Smart Building applications for several years [17]. Like BOT, SSN, SOSA, and SAREF, Brick uses OWL and RDF standards to define ontology and has an accessible website (https://brickschema.org/).

While each of the above mentioned ontologies could have been considered for this research, this paper focuses quantitatively on two – Haystack and Brick – that both represent pure versions of fundamentally different paradigms *and* have been the focus of industry R&D efforts. Because both its structure and semantic language are object-centered, Haystack – like IFC – is a pure example of an ontology using the object-centered paradigm, while the Brick ontology (described in detail in section 2.2) was selected to represent the description logic ontology language paradigm. Other ontological approaches were considered for the quantitative comparison, but these have not been indicated to be of interest to BAS vendors to incorporate into their systems. IFC, widely adopted by the AEC sector for BIM, could have been used to represent the object-centered paradigm; however, it is relatively unknown to industry as a potential ontology to support Smart Building applications. Additionally, while extremely complete with respect to equipment and location types, IFC 4.1 has only limited ability to represent the full range of BAS controls concepts [5], thus it is not an appropriate ontology for the detailed study. Because ifcOWL has been represented by an ontology language grounded in description logic (OWL), it does not represent a purely object-centered paradigm and could not be used for the comparison. Finally, the LBD groups BACS ontology was not use to represent a description logic ontology for comparison because it is only able to cover the domain of BAS semantics by integrating a number of other ontologies.

### 2.1 Project Haystack

Haystack [18] is a non-profit corporation, formed in 2014, and acts as the steward to the open-source Haystack Ontology. An industry board of directors and associate members maintain and develop the ontology. The board is comprised of individuals from smart edge hardware and software vendors including: ConserveIt, Intel, J2 Innovations (by Siemens), Legrand, Siemens, and SkyFoundry. Siemens is the major industry BAS provider involved in Haystack, though Honeywell is also indirectly involved through an associate member (Accutemp).

Haystack uses some terms exclusively to describe terminological and instance data. The Haystack ontology is a semantic data representation where dictionaries of *name value* pairs are defined; for example, to describe individual HVAC concepts in a building, where a *value* is the definition of the concept, and the *name* is the unique string representing the concept. The same *names* are used to describe instance data but are referred to as *Tags*. Name value pairs are called *Defs* [19], and are stored in portable groups called libraries; one or more libraries is used to define instance data [20]. Multiple *Defs* can be used to describe a concept and are referred to as *Conjuncts* [19]. *Defs* in the Haystack ontology can exhibit parent child relationships, effectively defining a hierarchy of concepts, referred to as *associations* [21]. Haystack instance relationships are defined using one of two methods: *Ref*, a type of *Def* that effectively functions as a pointer; or *Child Protos*, a *Conjunct* defined within a *Def*. Haystack relationships are derived from Haystack *relationship* types *containedBy*, *contains*, *receives*, or *supplies*. Haystack is serialized in Trio using the Zinc format; RDF serializations are not yet supported [22]. Implementing a Haystack instance is done in Trio, a plain text format, manually. Querying Haystack is done using a custom query language refers to as *Filters* [23]. **Error! Reference source not found.**Figure S1 shows a sample of Haystack describing an AHU bypass damper command point.

Earlier versions of Haystack consisted of a standardized vocabulary to define semantic concepts (i.e. a meta model). Haystack 4 has elevated its semantic representation to an ontology with hierarchy and relationships. Haystack is object-centered, attempting to digitally represent HVAC concepts in modular manner, as opposed to explicitly describing semantics. Haystack 4 documentation has stated that semantic web technologies would be supported, however this has not happened as of the version assessed in this research. A critique of Haystack has been its excessive flexibility, leading to instantiated ontologies that are not accessible to applications due to unexpected representations of semantics. The Haystack Tagging Ontology (HTO) proposed by Charpenay et al [14] is an application of Haystack which supports semantic web technologies, and structures the use of tags while extending the vocabulary included in the ontology. HTO uses Semantic Web technologies (RDF, OWL, SPARQL) to address this gap in Haystack [14].

## 2.2 Brick

Brick was initially described and used within the academic community,  initially published in 2016 [24], further publications have since added detail to the ontology description [5]. Brick is being collaborated on by researchers at Carnegie Mellon, Berkeley, University of California San Diego, University of California Los Angeles, University of Virginia, and the University of southern Denmark. Additionally, Brick is supported in industry by Johnson Controls and some regulatory bodies such as the US Department of Energy and the European Commission [25]. Haystack has more industry partners than Brick, however Brick offers more instantiated samples and more robust development documentation.

Unlike Haystack, Brick uses consistent language to describe teminologial and instance data. Brick was developed using a dataset developed by extracting data from six buildings using a variety of BAS vendors, and has been demonstrated to represent 98% of concepts [14]. Brick is grounded in description logic and was designed to describe building HVAC systems. Description logic "are a family of knowledge representation languages that can be used to represent the knowledge of an application domain in a structure and formally well-understood way" [26]. Brick exhibits a hierarchical design, where *Classes and Subclasses* define varying level of detail [14]. There are four primary Classes (Equipment, Location, Measurable, and Point). Additionally, there is a *relationship* Class that defines each of the nine bidirectional ontology relationships in subclasses. Brick uses a prescribed approach to defining semantics where complete HVAC concepts in an ordered string are represented in Classes and Subclasses. Figure S2 shows a sample of Brick representing an AHUs outdoor air damper command.

To facilitate the use of implemented Brick ontologies a query processor HodDB [27] has been designed. To achieve efficiency in query processing HodDB uses a decoupled database and query processor as specified by Fierro and Culler [27]. HodDB imports RDF serialized .ttl files containing building ontology data and stores it in a LevelDB, a database optimized for storing key-value pairs [28] . BTrDB [29] , optimized for the storage of timeseries data, is used to store timeseries data. In the HodDB architecture a UUID, unique identified associated to each semantic concept, allows the LevelDB and BTrDB databases to be cross referenced.

## 2.3 Ontology Comparison Methods

Ontology schemas are difficult to evaluate because they are declarative and only describe the domain that must be digitally represented rather than describing an instance of that domain. Evaluation through test cases often giving  quantitative results and is the preferred method to evaluate ontologies [2, 5, 14, 30]. Additionally domain-specific document comparison has been identified as a valuable means of ontology comparison [31]. Bajaj [32] reviewed the ontologies in the Smart Building domain based on their ability to respond to expected application queries, and support domain applications. It was found that ontologies that reuse existing ontologies, have been assessed with an ontology validator, are modular and accessible, are properly annotated, and well documented, are the most appropriate to support Smart Building applications. Many ontologies meet these criteria including Brick and Haystack which have been deemed in scope for this research.

Brick was validated using a case study method by Balaji et al. [5] . The ontology was compared to SAREF, Haystack and IFC, each being implemented for six buildings running eight Smart Building applications Each ontology *was measured in each building for completeness, expressiveness,* and *usability,* by assessing each ontologies recall of data points required by each of the applications. In this comparison Brick achieved higher completeness and expressiveness scores. Haystack has been validated using a case study by Bhattacharya et al. [33] comparing it against IFC and SSN using three buildings. Instead of using a set of Smart Building applications the Haystack validation used a summarized list of key relationships. The quantitative assessment done in the paper assessed *completeness,* and *expressiveness.* Additionally, f*lexibility was* discussed qualitatively. The paper found that Haystack offered the best completeness of the three assessed ontologies with 63% coverage of baseline data, as well as the best expressiveness with 77% coverage.

## 2.4 Summary of Literature Review

Many ontologies have been developed in the hopes of rendering building data sources, specifically BAS sources, more accessible to Smart Building applications. Despite the significant research to date on the topic of ontology to support Smart Building applications, there remains a paucity of literature comparing Brick and Haystack which use fundamentally different ontology language paradigms: the web based description logic approach and traditional object-centered approach, respectively. Brick and Haystack are recognized as viable representative ontologies for the applied comparison of paradigms . While previous versions of Brick and Haystack have been compared [5], each has been significantly updated during the intervening time, notably

Haystack, which has grown from a standardized schema to an ontology, and a new comparison for the purpose of Smart Building is required. This research aims to fill this gap by presenting a qualitative and quantitative (completeness and expressiveness) comparison of the Brick and Haystack ontologies.

## 3 Methodology

The study used the most current versions of both *target ontologies*: the beta Haystack Version 3.9.7 (referred to as Version 4 in marketing material) and its documentation as published on October 24th, 2019 [34], and Brick Version 1.1.0 and its documentation as published on February 21st, 2020 [35]. A comparative case study was undertaken using both quantitative and qualitative assessment to assess the ability of each of the target ontologies to accurately represent semantics of BAS systems for the purpose of Smart Building applications, considering both their *completeness* and their *expressiveness*. This process is referred to as *ontology validation* [30]. Ontology validation is the action of evaluating if the ontology matches the world model as demonstrated by the industry dataset, this is as opposed to ontology verification which employs tools to measure the quality of the ontology itself and whether it was properly built [30] . A case study approach was used, evaluating both ontologies against a large industry dataset called *KGS Clockwork.* This case study approach differs from previous research because it uses a *dataset* for comparison, described below, rather than one or more individual buildings. The completeness and expressiveness assessment results are interpreted relative to this dataset, to identify gaps in either target ontology. This approach offers the largest representation of building systems and data points and is therefore preferred to provide the most comprehensive comparative evaluation of Brick and Haystack.

### 3.1 Industry Dataset

KGS Clockworks is a for-profit organization providing customers a Smart Building application that provides equipment diagnostics, recommendations for energy savings, and occupant comfort condition ratings. Underpinning the KGS Smart Building application is a proprietary ontology representing BAS semantics. The offerings of KGS align with the goals of some Smart Building applications. The KGS ontology is not open-source, and therefore cannot be adopted by industry, however it is accessible to authors and has been known to represent the semantics of hundreds of buildings using BASs supplied by multiple vendors. Because this case study is focused on the comparison of industry ontologies to incorporate into BAS systems the KGS industry ontology, which has been successfully incorporated in a large number of commercial buildings, was selected as the industry dataset. This dataset is defined in a hierarchical structure, as seen in Figure 2. *Equipment Class* represents HVAC systems and common sub-equipment, while *Equipment Type* represents equipment subtypes. Equipment Types can be related to each other using an *Allowed Equipment Association*. Semantics of BAS data points are represented in *Point Types* which can be implicitly or explicitly related to equipment concepts. Point Type is a subclass of *Point Class,* which describe measures within a system. Each Point Class is associated with an *Engineering Unit* which defines the measurement of Point Classes. Finally, the *Measurement/Control Type* give DI/DO AI/AO context to Point Types, and *Service* indicates the medium the Point Types is acting on.

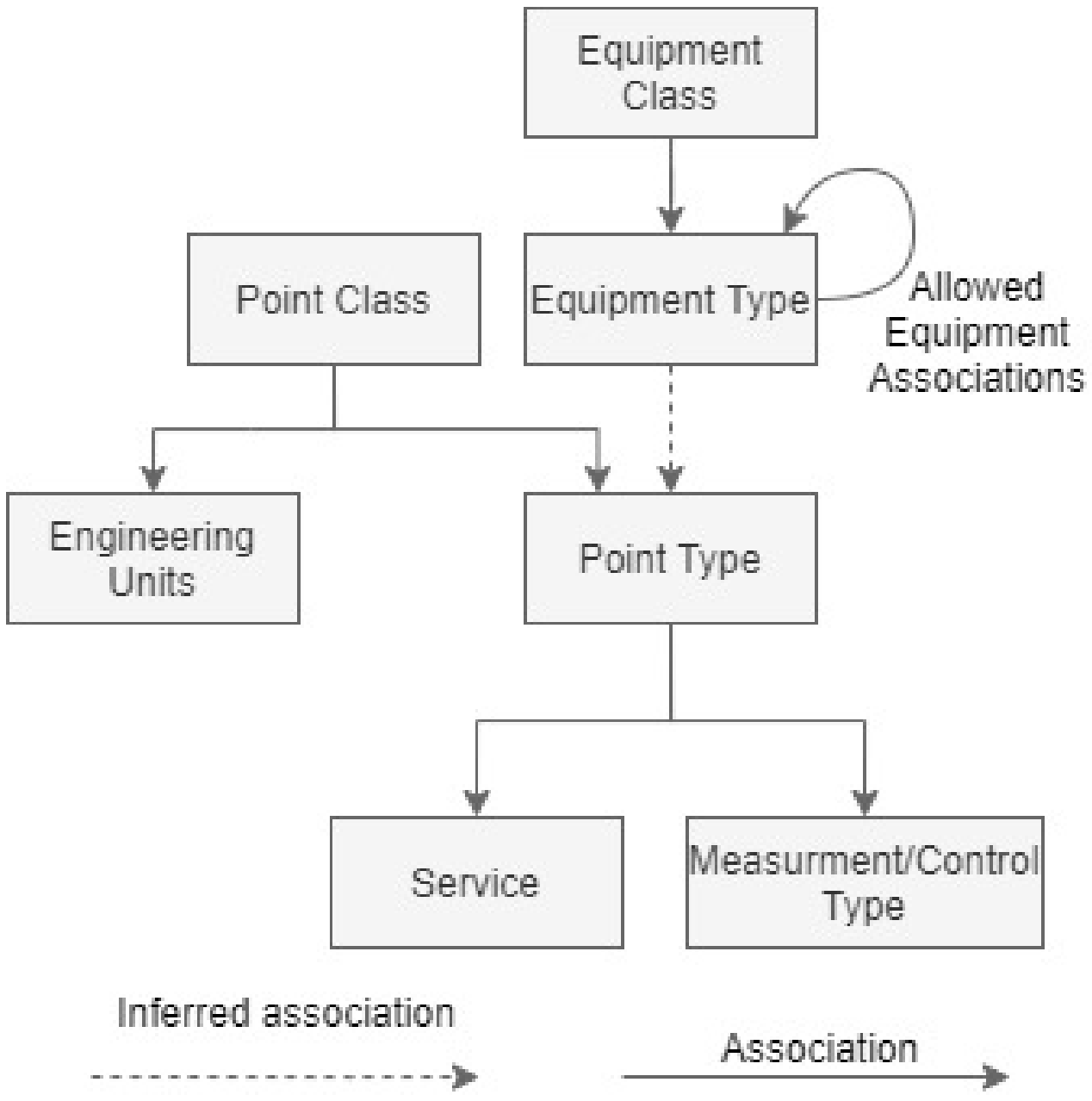

Figure 2: KGS Schema

The KGS ontology includes 1422 Point Types to capture the purpose of a data point in building systems. To manage the scope of *Point Types* included in the quantitative assessment, a subset of points associated with HVAC critical systems {AHU, Chiller, Boiler, Terminal Units, and Loops} were selected for mapping. Each critical systems redundant Point Types were removed, resulting in the *representative set* of 440 Point Types used for the comparative evaluation. Representative Point Types contain at least one unique *word* within the HVAC system that they are used where a word is a substring of the Point Type. For example, the point name *RoomAirDpTemp* was broken into the words: *Room*, *Air*, *Dp*, and *Temp*. This approach ensured that Point Types and their related equipment could be tested for representation in target ontologies. Some exceptions were made in selecting the representative set as not to include obscure Point Types used by KGS to represent "one of" system despite having unique words.

The breakdown of selected  and not selected representative Point Types by system is shown by subsystem in Figure 3 (Table S1 provides additional details on the selected points).

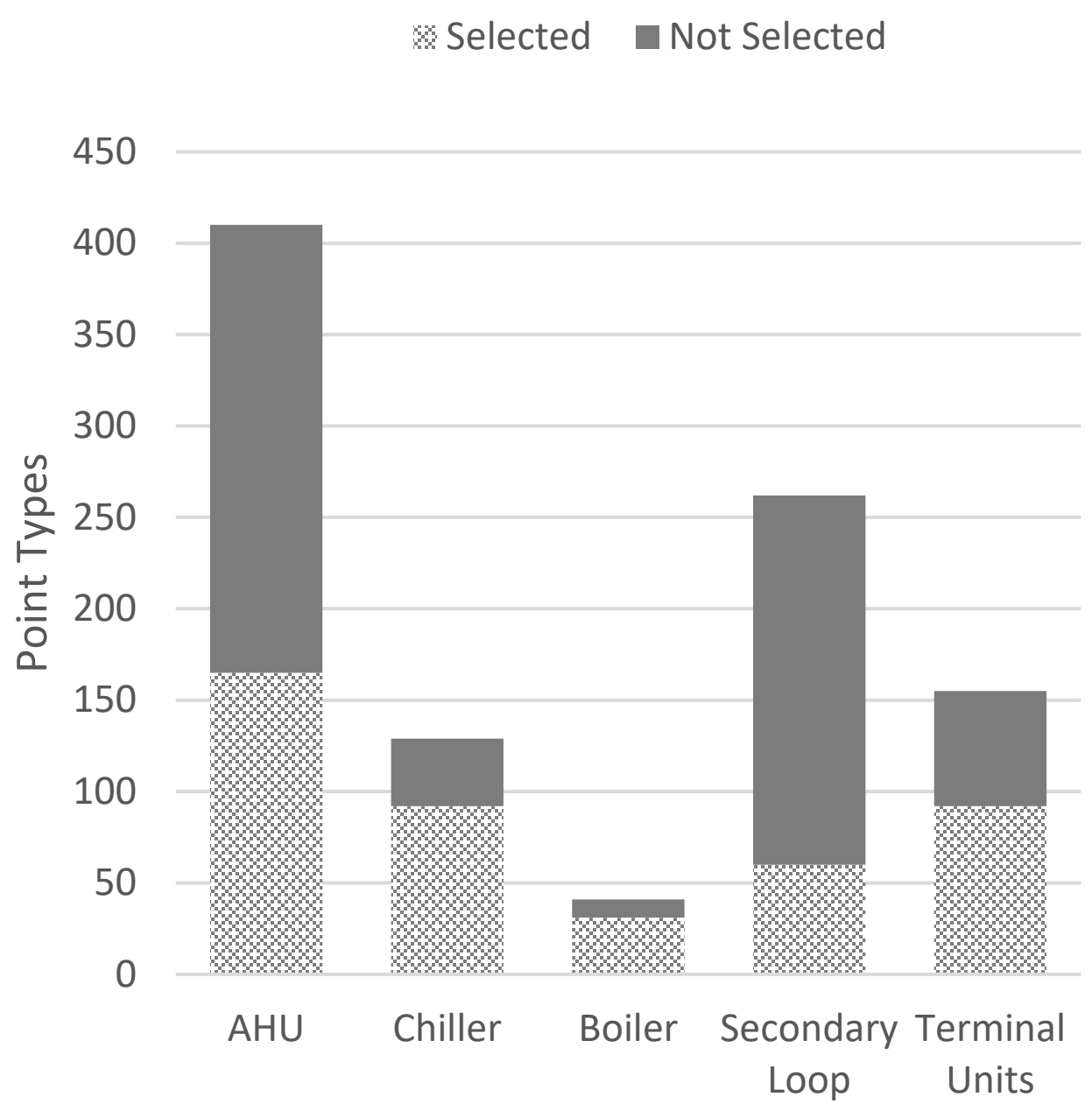

Figure 3 : Representative Industry Dataset Point Types by System

### 3.2 Quantitative Analysis

The quantitative assessment evaluates both completeness and expressiveness, which are measures demonstrated to be valuable in literature [5, 1, 27]. Completeness is a measure of the number of unique observations that can be made by an ontology [36] , and expressiveness is a measure of properties that can be expressed about observations [4]. Within the context of this research the measure of an ontologies completeness is the relative number of semantic BAS related concepts represented [5, 2], and expressiveness is the number of key smart building application required relationships expressed. To measure completeness and expressiveness a representative industry dataset was used as the baseline

#### 3.2.1 Completeness Measurement

Completeness was measured by assessing the percentage of representative Point Types that could be mapped from the industry dataset to each target ontology. Engineering Units were not considered in the completeness assessment as they were not easily identified for Point Types in the industry dataset; further, there are existing mature ontologies that represent the semantics of units the study of which is outside the scope of this research. Each representative Point Type was considered and classified when assessing the completeness of a target ontology. If *all semantic concepts* from the industry dataset Point Type could be mapped to the target the Point Type would be classified as one that *Maps.* If Equipment Class and Point Class mapped, but one gap existed in either measurement/Control Type, Service, or Equipment Type the Point Types was classified as *Partially Maps*. Finally, if there is more than one gap in Point Class, Service, or Equipment Type, or if the Equipment Class did not map, the Point Type was classified as *Does Not Map.* Examples of classifications of Point Types can be seen in Figure 4.

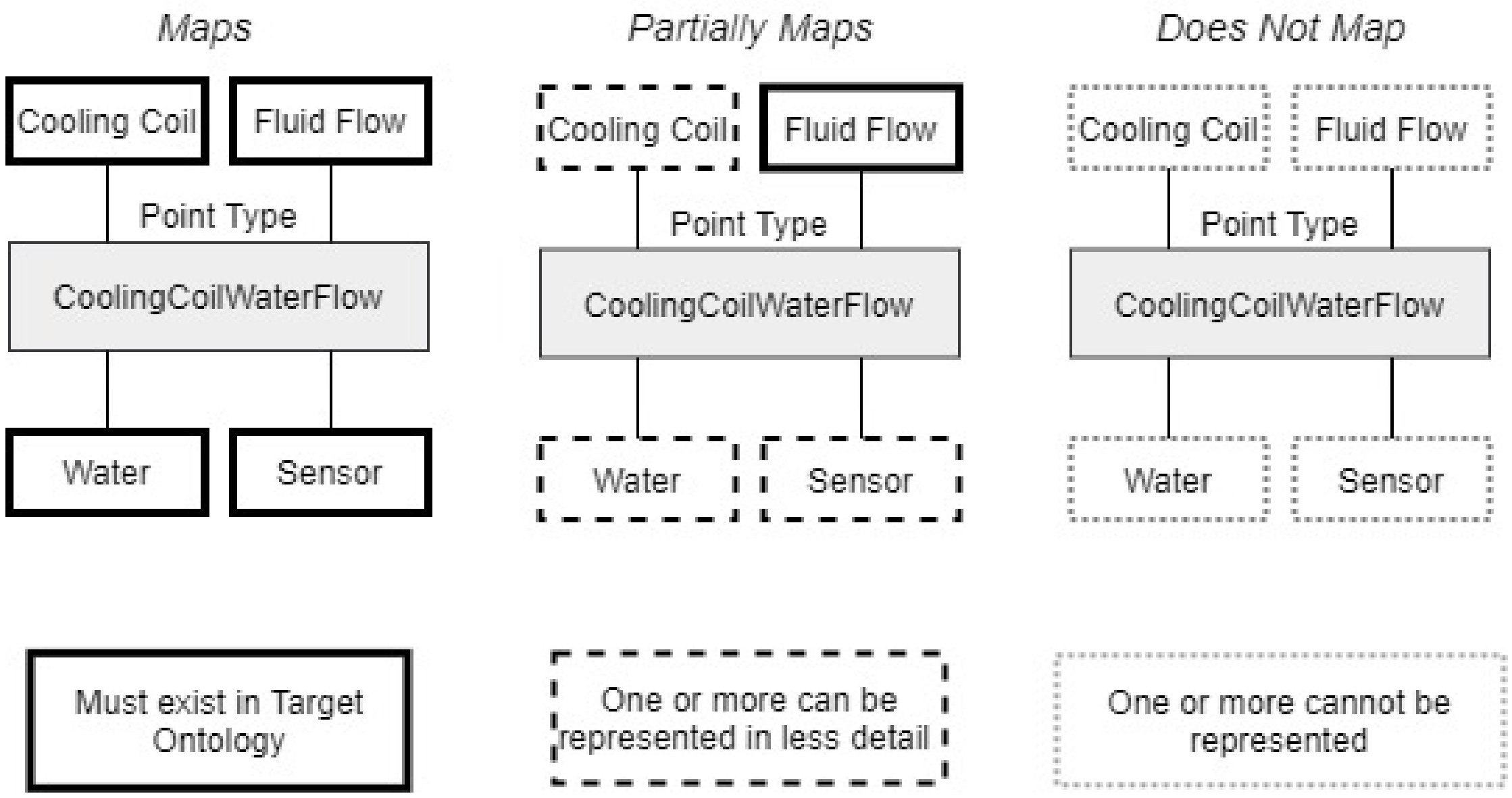

Figure 4: Target Ontology Point Type Classification Example

The use of word mapping semantic data from an industry dataset to a target ontology to asses completeness has been previously demonstrated within the literature [2, 5]. While the approach requires little manual effort, semantics beyond that explicitly stated in a word will not be evaluated. Direct word mapping semantic data was also not used in this research because the industry dataset Point Type word might not have an exact string match in the target ontology, despite having semantic mappings. For example, *HEX* in the industry dataset maps to *Heat_Exchanger* in the Brick ontology. This research used ontology alignment to manually map all connotations represented by words in each Point Type to ensure appropriate classification. Ontology alignment is defined as “given two ontologies each describing a set of discrete entities (which can be classes, properties, rules, predicates, etc.), find the relationships (e.g., equivalence or subsumption) holding between these entities” [37].

To better understand trends in completeness, *Partially Maps* and *Does Not Map* classified Point

Types were assessed for gaps. Each piece of missing information (a gap) in a Point Type was coded as a measure, equipment, medium, or concept, and the missing concept was indicated. Point Type could thus exhibit multiple gaps, and each would be considered in the completeness trend analysis.  Gaps were interpreted relative to their *Does Not Map* and *Partially Maps* classification and were classified as either significant (affecting 2% of more of Point Types in the representative set) or insignificant.

### 3.2.2 **Expressiveness Measurement**

Expressiveness was measured by quantifying a set of key relationships. Each of these key relationships was found in the industry dataset for the set of previously identified systems (AHU, Chiller, Boiler, Terminal Units, and Loops) on both the air and water sides. The total number of relationships assessed was 27.

*Key relationships* required by Smart Building applications have been identified in previous publications [33, 5, 27, 1]. These include the six relationships {Sensor ↔ Location, Location ↔ Location, Equipment ↔ Location, Sensor ↔ Equipment, Equipment ↔ Equipment, Location ↔ Persons} posited by Bhattacharya et al. [2] and one additional relationship {Equipment ↔ Name} based on Balaji et al. [5]'s expanded relationship set used for ontology validation.   These key relationships were also used in research for query processors serving Smart Building applications [1] [27]. The industry dataset defines semantic relationships within Point Types and Allowed Equipment Associations. Key relationships were cross-referenced with the industry dataset to guide the selection of the set of representative industry dataset *expressed key* *relationships* (Table S2). Using a similar logic to completeness, trends in expressiveness were assessed by classifying expressed key relationships as *Maps* or *Does not Map*.

### 3.3 Qualitative Comparison

A qualitative assessment was performed for each target ontology along with a breadth of other ontologies to support a critical analysis of the strengths and weaknesses of Haystack and Brick within the broader ontological landscape. To undertake this evaluation, the ontology documentation was reviewed, assessing the flexibility, portability, readability, extensibility, interoperability, and queryability of each ontology, using a set of competency questions.

The *qualities* mentioned above have been regarded as positive and relevant in the academic literature [2, 5, 38, 39]. *Flexibility* answers the question 'can the ontology capture uncertainty" and "does it use non-restrictive methods to define concept semantically?' [2]. *Portability* answers the questions 'can the same set of applications be applied across buildings (with applicable HVAC systems) using the specified ontology?' and 'are concepts represented consistently in a machine-readable format.' [5]. *Readability* answers the question 'can domain experts and applications developers unambiguously decipher real world meaning from semantics as presented in the ontology?' [5]. *Extensibility* answers the question 'can the ontology be customized to add new semantic concepts?' [2, 5]. *Interoperability* answers the questions 'can the ontology integrate with, and convert to, other ontologies with little to no human effort?' and 'Is the ontology serialized in an industry accepted format?' [38]. Finally, *Queryability* answers the questions 'can an instantiated ontology be machine traversed and necessary information retrieved?’ and ‘Is there low variability in semantic relationships?’ [39, 33]. The response of yes is the desired outcome to quality questions, indicating that the quality is expressed in the ontology.

# 4 Results

The following results present the comparison of completeness, expressiveness, and qualitative assessments of Brick and Haystack. Completeness and expressiveness assessments used the KGS industry dataset ; as noted previously, this ontology was developed for Smart Building applications that performed fault detection, energy optimization, and human comfort tracking functionality, and is therefore well suited to provide relative results. Completeness was measured by quantifying the number of industry dataset Point Types classified as *Maps* in the target ontology while expressiveness was measured by quantifying the number of industry dataset expressed key relationships classified as *Maps* when mapping comparable relationships from the industry dataset to the target ontology. Finally, a defined set of desirable qualities were assessed for representation in either target ontology by evaluating relevant documentation and classifying qualities as *supported* or *unsupported* these qualitative results were contextualized by comparing the two ontologies to others relevant to the domain.

## 4.1 Completeness

Brick was found to be more complete than Haystack with a higher percent of representative Point Types classified as Maps: 59% vs. 43%. When Point Types classified as Map or Partially Maps Brick again achieves higher completeness with 77% of Point Types covered, while only and 69% of Point Types were so classified for Haystack. Brick was able to represent a greater number of representative Point Types from the industry dataset than Haystack. This was true across all subsystem considering Point Types classified as Maps, excluding the Boiler system which achieved the same value; this is illustrated in Figure 5.

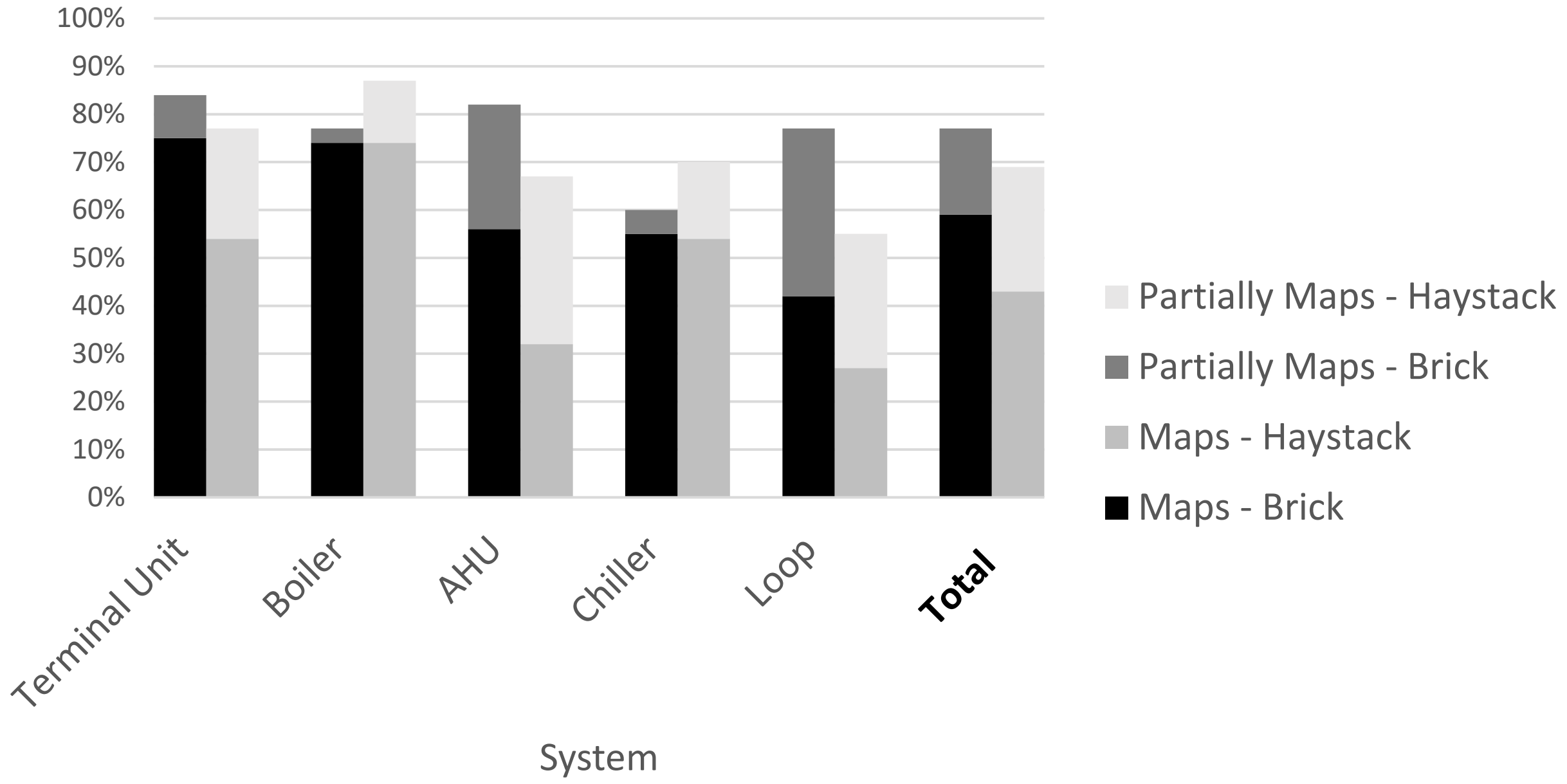

Figure 5 : Brick and Haystack Completeness Results

The numerical results for the completeness of Haystack and Brick by system type can be seen in Table 1. Brick offered a more complete ontology where a greater number of semantic concepts were represented. There were fewer Significant gaps in Brick and ultimately it would require fewer custom semantics to be defined to accurately represent a building.

Table 1 : Completeness Results Against Industry Representative Point Type Set

| System | Haystack | | Brick | |
|---|---|---|---|---|
| | % Point Types with Maps Classification | % Point Types with Maps or Partially Maps Classification | % Point Types with Maps Classification | % Point Types with Maps or Partially Maps Classification |
| AHU | 32% | 67% | 56% | 82% |
| Chiller | 54% | 70% | 55% | 60% |
| Boiler | 74% | 87% | 74% | 77% |
| Loop | 27% | 55% | 42% | 77% |
| Terminal Units | 54% | 77% | 75% | 84% |
| Total | 43% | 69% | 59% | 77% |

A primary finding of completeness measurement was the identification of semantic gaps in either target ontology. This study found the Haystack ontology had more (60) unique gaps and overall occurrences (303) than Brick (50 gaps and 208 occurrences). These gaps are shown in

Table 2 (Haystack) and

Table 3 (Brick). Significant Gaps (those affecting 2% or more of Point Types in the industry dataset) are discussed below. Haystack had 8 Significant gaps, where Brick has 6. There are many Insignificant Gaps (those affecting less than 2%) which, in aggregate, have a large impact on completeness; each impact only a few Point Types but are nonetheless necessary in the industry dataset. It should be noted that these Insignificant Gaps could be more significant if the representative Point Type set were expanded to include all Point Types in the industry dataset. Gaps could be filled by extending either ontology following its ontology specification paradigm. Where this research considers ontologies for BAS data gaps specific to this domain. Gaps outside of this domain might be better representing by an integrating existing ontologies to avoid a monolithic ontology that cannot be easily queried similar to the method BOT has taken in representing buildings semantics with many OWL ontologies.

Table 2 : Semantic Gaps of Haystack Ontology Affecting Completeness Assessment Score

| ***Gap Type*** | ***Significant*** | ***Classification*** | ***Gap (# Point Types)*** |
|---|---|---|---|
| Missing Concept | Yes | Does Not Map | Alarm (13) |
| | Yes | Partially Maps | Primary/ Secondary (24) |
| | No | Does Not Map | Conditioning Mode (1), Pre Heat (1), Set Back Status (1), Setup Mode (1), Relief (1), Holiday (1), Superheating (1), Natural Ventilation (1), Occupancy Override (1), Part Run More (2), Tracking Mode (2), Subcooling (2), All Run Model (2), Loop Overlap (2), Runtime (3), Tracking Status (3), Free Cooling (4), Part Load (6) |
| | No | Partially Maps | Heat Source (1), Medium (3), Low (5), High (6), Reset (8) |
| Missing Equipment | Yes | Does Not Map | Heat Recovery (15) |
| | Yes | Partially Maps | Generic Compressor (18), Enthalpy Wheel (20) |
| | No | Does Not Map | Hot Water Loop (2), Humidifier (3), Economizer (7), Thermal Energy Storage (3), Generator (4), Generator (4), Filter (5), Dual Temp Loop (7) |
| | No | Partially Maps | Dual Temp Coil (3) |
| Missing Measure | Yes | Does Not Map | Enthalpy (9) |
| | Yes | Partially Maps | Position (54) |
| | No | Does Not Map | Vibration Amplitude (1), Volume (1), Suction Pressure (1), Oxygen Fraction (1), Ph (1), Cooling Rate (1), Fire Rate (1), Heating Rate (2), Humidity Ratio (2), Boiling Temp (2), Conductivity (2), Velocity Pressure (3), Static Pressure (6) |
| | No | Partially Maps | N/A |
| Missing Medium | Yes | Does Not Map | N/A |
| | Yes | Partially Maps | Equipment Discharge Air (10) |
| | No | Does Not Map | CO (2), Lubrication Oil (3), Return Water (3), Supply Water (4) |
| | No | Partially Maps | Clean Steam (2), Equipment Inlet Air (6) |

Table 3 : Semantic Gaps of Brick Ontology Affecting Completeness Assessment Score

| ***Gap Type*** | ***Significant*** | ***Classification*** | ***Gap (# Point Types)*** |
|---|---|---|---|
| Missing Concept | Yes | Does Not Map | N/A |
| | Yes | Partially Maps | Setpoint Limit (21), Primary/ Secondary (22) |
| | No | Does Not Map | Fan Only (1), Superheating (1), Cooling Enable (Outdoor Air ) (1), Setback Status (1), Setup Mode (1), Holiday (1), Part Run (2), All Run (2), Tracking Mode (2), Subcooling (2),Tracking Status (3), Stage Command (3), Free Cooling (4), Heat Source (5) |
| | No | Partially Maps | Occupancy Mode (1), Return Air Reset (4) |
| Missing Equipment | Yes | Does Not Map | Heat Recovery (15) |
| | Yes | Partially Maps | Enthalpy Wheel (20) |
| | No | Does Not Map | Mixing Valve (1), Relief Damper (1), Thermal Energy Storage (3), Generator (4), Pre Heat/Cool Coil (5), Radiant Terminal Unit (6) |
| | No | Partially Maps | Dual Temp loop (1), Face Damper |
| | | | (1), Cold Deck (3), Hot Deck (3), Bypass Valve (7) |
| Missing Measure | Yes | Does Not Map | N/A |
| | Yes | Partially Maps | N/A |
| | No | Does Not Map | Fire Rate (1), Oxygen Fraction (1), Ph(1), Suction Pressure (1), Vibration Amplitude (1) Illuminance (1), Cooling Rate (1), Heating Rate (2), Humidity Ration (2), Boiling Temp (2), Efficiency (2), Differential Pressure (3) |
| | No | Partially Maps | Volume (2) |
| Missing Medium | Yes | Does Not Map | Refrigerant (21) |
| | Yes | Partially Maps | Equipment Inlet Air (9) |
| | No | Does Not Map | Clean Steam (2), CO (2), Process Water (6) |
| | No | Partially Maps | Equipment Discharge Air (1) |

Figure 6 offers a visual representation of these individual and shared gaps classified as Does Not Map. Haystack exhibited eight such Significant gaps, six being unique and included the lack of alarms and generic compressors, additionally it could not represent the measurement of enthalpy, position, and sub equipment discharge air. Three of Haystack's Significant gaps overlapped with those in Brick and included the lack of description for enthalpy wheel and heat recovery equipment, as well as the ability to differentiate between primary and secondary equipment. Brick exhibited six Significant gaps, three as previously stated and three of which were unique and included the lack of a refrigerant substance, sub-equipment inlet air, and limits. On a system level, loops were the most poorly represented, with the lowest completeness scores. Smart Building applications will benefit from the higher completeness offered by Brick as a wider breadth of HVAC concepts can be described and therefore accessible by applications. Buildings using a more complete ontology could be compatible with a wider variety of Smart Building applications, specifically those relying on clear and descriptive semantics. While gaps in ontology schema can be filled as to not affect application effectiveness, their completion is preferred to avoid additional work in defining semantic concepts when developing a Smart Building application.

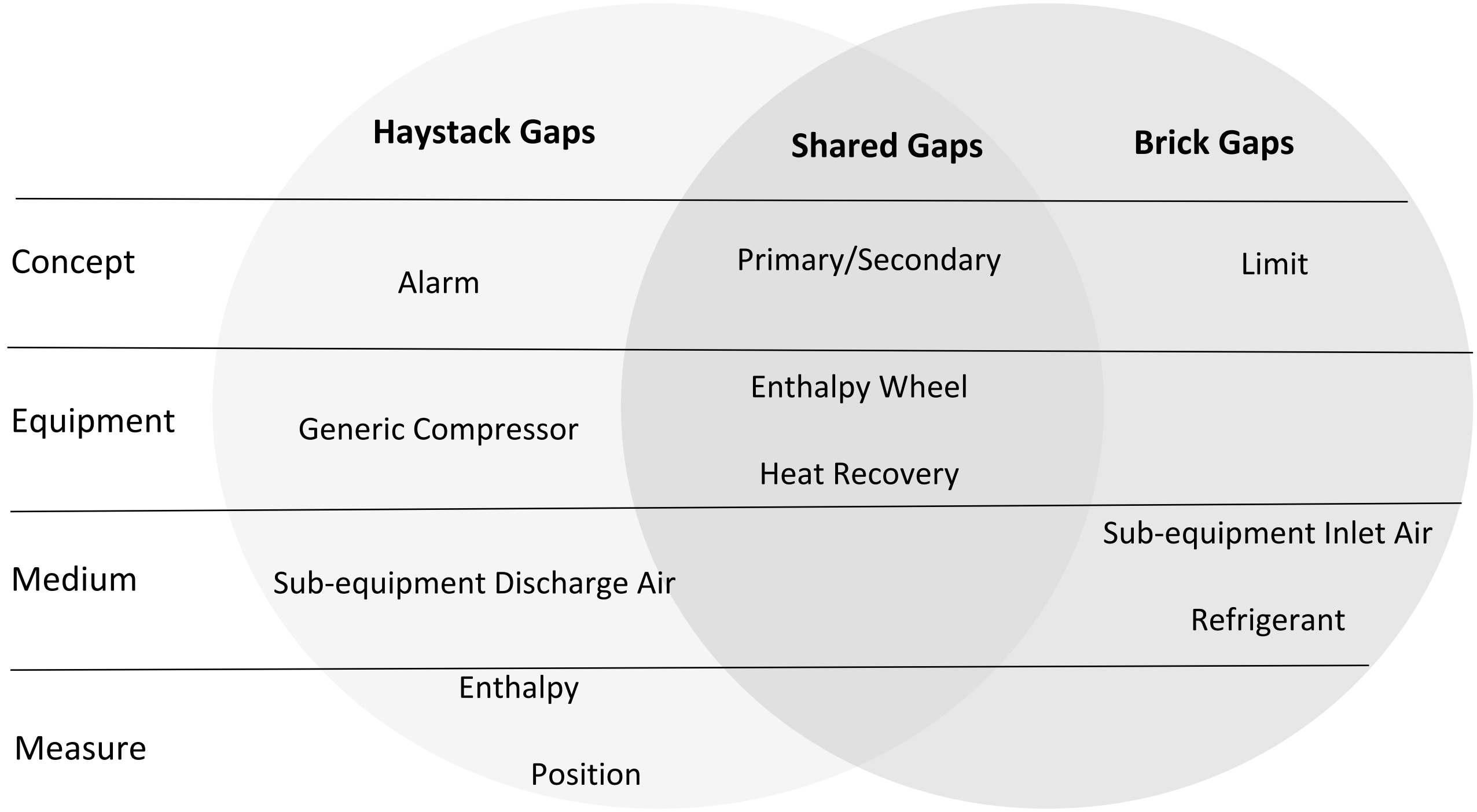

Figure 6 : Overlaps and Discrepancies in Brick and Haystack Ontology Significant Gaps classified as “Does Not Map”

### 4.2 Expressiveness

As noted in the methodology, expressiveness was measured by quantifying the number of key relationships required by Smart Building applications as found in the industry dataset that map to the target ontology. The Brick ontology was found to be marginally more expressive than Haystack. Brick relationships are well suited for describing representative industry dataset expressed key relationships, 100% of the 27 relationships assessed were classified as Maps. Haystack relationships were able to describe almost all representative industry dataset expressed key relationships: 96% were classified as Maps, all but one sub-equipment relationship. The Brick ontology explicitly defines relationship function and their constraints whereas the relationship function in Haystack is implicit. The explicit and implicit approach to relationships of Brick and Haystack align with eithers overall ontology schema where Brick is prescribed, and Haystack more flexible.

A common relationship used for Sensor ↔ Equipment relationship mapping was *equipRef*, this relationship is used to reference an equipment, which contains the sensor appropriate tag being used. Only ten other *Ref* relationships are defined in the Haystack ontology, of which *ahuRef*, *hotWaterPlantRef*, and *chilledWaterPlantRef* were used in mapping. In addition to *Ref* relationships, *Defs* can have *Child Protos* defined and are *contained* by a *Def*. *Child Protos* were used to map the Equipment ↔ Equipment air side relationship. Haystack relationships were able to bridge the gap between the air and water side of HVAC systems; however, this required the use of both *Ref* and *Child Proto* relationships which in practice would necessitate complex queries.

The set of Brick relationships is clearly defined in a *Class* with defined constraints (Table S3). Brick has nine bidirectional relationships, each with a defined inverse relationship. The explicit support of direct inverse relationships made expressiveness assessment simpler as only one use/direction of the relationship needed to be found to confirm the support of the bidirectional key relationship. For example, it was found that the Brick relationship *hasPoint* can be used to represent the Boiler Equipment → Sensor relationship, therefore it is known that the inverse relationship *isPointOf* can be used to represent the Boiler Sensor → Equipment relationship. The most used Brick relationships in the expressiveness assessment were *hasPoint/isPointOf*, *hasPart/isPartOf*, and *feeds/IsFeedBy*. These relationships were used because their constraints aligned with end points in key relationships such as Equipment and Sensor. Other Brick relationships such as *measures* and *regulates* are better suited to relate more granular semantic concepts such as measurables.

A single Brick relationship could not describe some Equipment ↔ Equipment relationships, specifically those between loops and other HVAC systems. These relationships could be represented in Brick using the *feeds/isFedBy* relationship; however, the *feeds* relationship is permitted only for a sequential process where a media is passed between the two end points. Because a loop is a cycle and the media within the loop changes as it interacts with different equipment, multiple relationships are needed to be used to represent their key relationships. For example, the Chiller↔ Loop could be represented with Chiller *feeds*→ Loop and Loop *feeds*→ Chiller. Where the first relationship is passing chilled water from the chiller to the loop, and the second is passing return warm water from the loop to the chiller.

### 4.3 Qualitative Analysis

The qualitative analysis assessed the target ontologies for six desirable qualities by responding to answering competency questions given statements made in supporting documentation, results can be seen in Table 4, and are discussed in more detail below. Brick exhibited five of the six, where Haystack only exhibited three. The Brick schema is based in description logic whereas Haystack is object-centered. These fundamentally different ontology language paradigms affect Linked Data approaches and the qualities the ontologies can exhibit.

Table 4: Qualitative Results

| ***Quality*** | ***Competency Questions*** | ***Brick*** | ***Haystack*** |
|---|---|---|---|
| Flexibility | Can the ontology capture uncertainty and does it use non-restrictive methods to define concept semantically? | | ✓ |
| Portability | Are semantics represented consistently in a machine-readable format that is building agnostic? | ✓ | |
| Readability | Can domain experts and applications developers unambiguously decipher real world meaning from semantics as presented in the ontology? | ✓ | |
| Extensibility | Can the ontology be customized to add new semantic concepts? | ✓ | ✓ |
| Interoperability | Can the ontology integrate with, and convert to, other ontologies using an industry accepted format with little to no human effort? | ✓ | |
| Queryability | Can the instantiated ontology be machine traversed and necessary information retrieved? Is there low variability in semantic relationships? | ✔ | ✔ |

### 4.3.1 **Flexibility**

Haystack offers flexibility through the *Tag* based schema using *Defs*, which focuses on representing smaller units of semantic information than Brick. *Tags* allow Haystack to represent uncertainty by allowing a limited number of *Tags* to be used when representing minimal semantic information. The freedom of using small units of semantic information to represent concepts offers a non-restricted way of defining semantic data, whereas Brick ensures concepts are prescribed with their more descriptive *Classes* composed of a set of unit *Tags*. The flexibility exhibited by Haystack will allow the ontology to represent a wide variety of buildings typologies with HVAC systems that might fall outside of the norm. Flexibility will also allow for the semantic description of a subset of specific concepts within a building if the whole ontology is not desirable and beneficial to a Smart Building application. Flexibility could decrease the portability of Smart Building applications across buildings.

For example, within Haystack the Point Tag has a non-mandatory marker pointFunction, which includes the choice of -cmd (command), sp (setpoint), or sensor (data value reading). This marker could be used to map the industry dataset Measurement/Control Type value of all Point Types. Alternatively, these values could be tagged with one of the Point subtypes cur-point, his-point, weather-point, or writable-point. These subtypes offer the option to define more detailed semantic data. Industry dataset Point Types mapped to point *Tags* with defined *pointFunction* markers would be retrieved by a SOCx application using different queries than Point Types mapped to the appropriate Haystack *Point* subclass *Tags*, which do not have an assigned *pointFunction* since is not mandatory marker.

The BOT approach to ontology encourages flexibility through the modularization of concepts using separate existing ontologies that each represent a specific set of building related concepts. For example, Building Automation and Control System ontology (BACS) is used to represent BAS semantics and the Sensor, Observation, Sample, and Actuator (SOSA) ontology is used to represent more specific sensor concepts such a “features of interest” not covered in BACS. Alternatively, IFC approaches flexibility through generalizations such as IfcPropertySetDefinition, which can be either statically defined, i.e. already in the IFC schema, or dynamically extendable. In the latter case, there is no entity in the "meta model"; instead the value is declared by “assigning a significant string value to the Name attribute of the entity as defined in the entity IfcPropertySet and at each subtype of IfcProperty, referenced by the property set” [40]. The IfcPropertySetTemplate provides a means for creating such new property sets such that they remain consistent with IFC 4.

#### 4.3.2 **Portability**

The Brick schema’s prescribed representation of whole HVAC concepts ensures consistency across ontology instantiations facilitating Smart Building application portability. The same Smart Building application can be used in multiple buildings that employ Brick because similar HVAC system components across building are guaranteed to use the same Classes or subclasses of those. Alternatively, Haystack’s *Tag* based schema does not ensure consistency because it is flexible. As exemplified in Section 4.3.1 Haystack's flexibility can result in BAS concepts being represented inconsistently in instantiations; this quality does not ensure that standard portable queries could be written and used. Both ontologies are machine readable because they are serialized in a standardized format.

The LBD groups BOT approach includes ontologies with overlapping concepts such as space illuminance, defined in BACS as *SpaceIlluminanceObs*, and IFC as *ifcIlluminanceMeasure*. The BASC ontology defined specifically for BAS ontology within BOT uses *Unions* to remediate this issue, where a union defined equivalences of ontology definitions. However, the current specification [http;//www.ontoeng.com/bacs#] does not contain many Unions and more would be needed to ensure the portability of Smart Building applications.  Alternatively, IFC has proven it as portability through the widespread adoption of the standard in the AEC industry. IFC, a single schema, can undeniably represent domain data for AEC, and offers proof that a single standard - if accepted within industry - can be used to facilitate portability of information across a variety of applications.

### 4.3.3 **Readability**

Brick was designed using description logic and is therefore a readable ontology with explicitly defined semantic concepts and relationships. Semantic concepts are represented with a hierarchy of Classes: Equipment, Point, Measurable, Relationships, and Location, setting a clear expectation of ontology use for end users. Alternatively, Haystack's use of *Tags* can be used to represent semantic concepts in a variety of ways. The potential for inconsistency in semantic representation yields an ontology that is difficult to read and relate to real world meaning with confidence. This will be especially true for concepts defined with fewer tag in Haystack when more semantic description might be required to differentiate BAS concepts. An ontology that is not human readable will mean that end user could have difficulty querying the ontology to access time series building HVAC system data not directly available in the BAS. The complexity and integration of multiple existing ontologies might negatively impact the readability of an instantiated ontology, requiring a user to comprehend several hierarchical ontologies and their interaction. This would be true for the LBD group's approach which integrates BOT, BACS, and SSN, among others.  While IFC is a single a well-known industry standard, it could be difficult for experts and developers to unambiguously decipher the real-world meaning of concepts because of the vast number of spatial descriptions in the ontology. While many of these spatial structural concepts are not used to describe BAS data, the definition of BAS data within the existing complex instantiation of BIM data could be overly complex and impact readability. All of that said, IFC meets the requirements of unambiguously representing concepts.

#### 4.3.4 **Extensibility**

Both Haystack and Brick ontologies are extensible. The Brick can be extended by updating the schema definition file. Concepts are added to Brick by: (1) naming the concept, (2) placing it within the existing class hierarchy, and (3) defining tags associated with the concept. This is achieved using a Python script used to build the schema definition (.ttl) file. Brick extensions can be incorporated through a git repository following standard git contribution practices where the contributing community will vet suggested extensions [41]. Haystack can be extended by defining new *Defs* directly in a library (.trio) file following the custom schema syntax (Zinc). Contributors of Haystack are encouraged to participate in the ontology forum; however, Haystack also has an open source git repository where Defs are defined and users could issue a pull request to make their own extensions. Other ontologies, including BACS also used this open source repository approach. The formal governance structures dictating the valuable extensions that will be added to future version of Brick and BOT ontologies are ultimately made by academic contributors who are owners of the repositories. It should be noted that ontologies can be extended using different approaches, for example BOT is a small central ontology that is meant to be extended by aligning domain ontologies domain ontologies can be aligned rather than directly extending. IFC offers a more formal governance model, where the standard is recognized by an ISO committee, and maintained by the buildingSmart Model Support Group. Further, the presence of the IfcPropertySetTemplate facilitates such extension.

#### 4.3.5 **Interoperability**

The Brick ontology could be interoperable with other ontologies serialized in the RDF format (e.g. BOT and SAREF) using the Linked Data approach described by W3C [38]. The BOT definition has in fact leveraged RDF and provides an initial alignment of the small central ontology to the Brick ontology for the description of BAS semantics. Ontologies supporting the RDF format can be integrated with Brick by manually cross-referencing common concepts and stating their Unions within the ontology definition, giving Smart Building applications robust access to data. This is true for all the description logic ontologies defined using the W3C standard, however it should be recognized that this is a difficult and manual process and some definitions between ontologies may not align perfectly, human negotiation of concepts is necessary to properly integrate ontologies. IFC could also be integrated in this way using its OWL representation (ifcOWL). Brick take interoperability a step further and defines tags related to each Class which can be used to directly convert an instantiated Brick ontology to Haystack making the Brick ontology interoperable with Haystack. Haystack is currently in the process of building functionality to serialize to the RDF format; however, it is not yet supported.

#### 4.3.6 **Queryablility**

Brick uses a small set of nine bidirectional relationships and relies on SPARQL, this facilitates the consistent retrieval of semantic data given the limited number of relationships and the high capacity for complexity in queries. IFC, represented in the EXPRESS language, lacks some knowledge representation found in description logic based ontologies [42] and a generally accepted query engine. As a result, there was the demand to represent IFC using OWL [43]. Brick also relies on the HodDB [1] to efficiently retrieve timeseries data, a database specifically designed to query timeseries data via an RDF ontology. Ontologies supporting W3C technologies could conceivably use in this database system.  Haystack documentation describes a query method ("Filters") that allows for the retrieval of semantic data using basic logic. Haystack queries, although logically simpler than SPARQL, can be mapped to JSON which through APIs opens the door to a plethora of timeseries optimize database solutions. Queryability affects accurate data recall, Smart Building applications would not be able to access timeseries building data without accurate data recall.

## 5 Conclusions and Discussion

Of Brick and Haystack, Brick is a better-suited ontology to represent and relate HVAC concepts for use by Smart Building applications because it is more complete (77% of points fully or partially map), expressive (100% of relationships can be mapped), and is superior from a qualitative assessment. This said, Brick still lacks several concepts and is thus incomplete. Fortunately, several ontologies exist that can fill these conceptual gaps, notably those developed using W3C (BOT, SAREF, SSO, BACS). Because Brick is grounded in description logic and uses a communicative manner that is clear and concise to represent BAS concepts, it can be integrated with other W3C AEC ontologies through human negotiations and effectively support Smart Building applications. While Haystack achieves comparable scores in completeness (69%) and expressiveness (96%), it was only able to achieve three of the six key qualities based on the competency questions from the literature, while Brick achieved five. Further, Haystack cannot easily be integrated with other AEC ontologies due to the ontology language used to define it.

The Brick schema was structured in a prescribed manner that allows for the qualities of readability and portability to be supported, which were not supported by Haystack. These qualities facilitate Smart Building application development because concepts are represented unambiguously and consistently. Portability and readability together allow Smart Building applications to be modular, where they can be implemented across multiple buildings. IFC has previously demonstrated the value of portable data standards in industry [44]. Bricks use of W3C semantic web technologies, including RDF and SPARQL, facilitate the support of interoperability and queryability opening the door to Brick federation with other ontologies to fill schema gaps. The interoperability and queryability of Brick will allow it to integrate with other ontologies and ultimately provide Smart Building applications the most robust and holistic representation of building data. Brick's support of SPARQL, a standardized query tool, will allow Smart Building applications to query the data, however Haystack exhibits an adaptable query method compatible with JSON. Flexibility not exhibited by Brick can be accommodated by its integration with other W3C ontologies, such as BOT.

Each of the SAREF [https://ontology.tno.nl/saref.ttl], BACS [http://www.ontoeng.com/bacs#], Brick [https://brickschema.org/ontology/1.1], and IFC [https://standards.buildingsmart.org/IFC/DEV/IFC4_3/RC1/EXPRESS/IFC4x3_RC1.exp] schemas were reviewed to understand how Brick could contribute to the ecosystem of ontologies representing BAS concepts in Figure 7. IFC covers the most relevant concepts and should be considered as a starting point. As discussed in section 4.3, the set of ontologies using W3C technologies have strengths and weaknesses but would ultimately allow for all building data to be semantically represented. The integration of multiple standards will require careful human negotiations. Additionally, such a standard would benefit from a formalized governance approach as provided by buildingSmart for IFC.

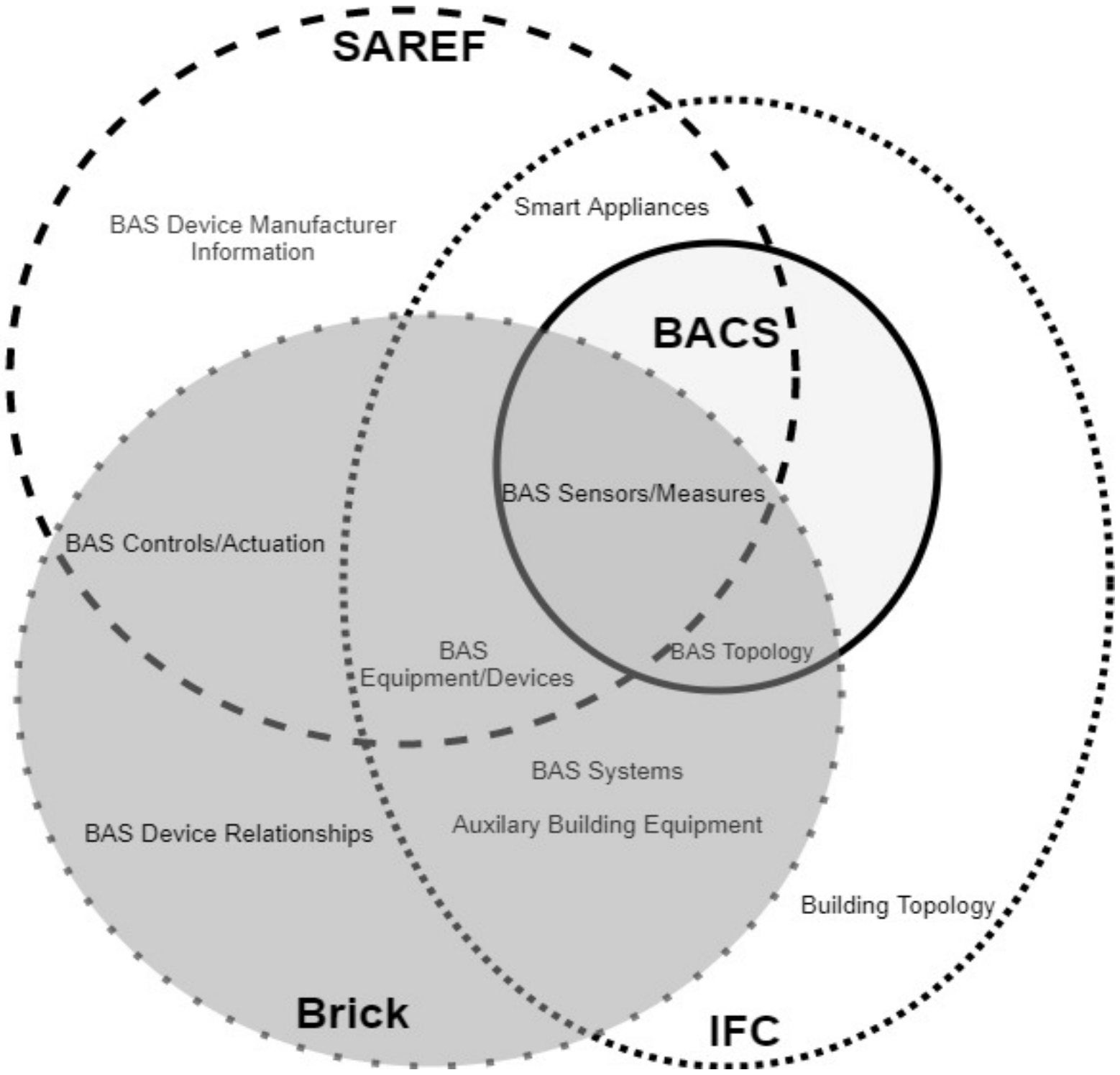

Figure 7: W3C based ontologies

Although the expressiveness assessment was practically inconclusive in distinguishing between Brick and Haystack ontologies with respect to relationship representation, it informed further consideration of the *quality* of relationships in target ontologies. IFC's inability to represent some knowledge representation due to its use of the object-centered EXPRESS language is a testament to the need for a BAS ontology which support the W3C standards. Although ifcOWL is being developed to support these standards, the ontology could benefit from BAS device relationship and the control/actuation concepts defined in Brick. Haystack used two bidirectional relationships to create Refs and Child Protos that relate various concepts but otherwise added no semantic information. Alternatively, Brick used a set of nine descriptive bidirectional relationships, each describing a type relationship used in a specific scenario. While Haystack relationships achieved a high score in expressiveness, they were not descriptive and as such the nature of the relationship could only be inferred. Smart Building applications will benefit from the small but descriptive set of relationships used in Brick because they offer more semantic information not provided in Haystack relationships.

Ontology comparisons usually use one or more buildings as test cases, and in every such study, the content of the test inputs impacts the results. Because it has synthesized the data from hundreds of buildings, the industry dataset was assumed to be complete and accurate it's representation of BAS concepts, at least as of the time of this publication. Like all studies using real-world data, however, any missing points or relationships – whether omitted or simply concepts implemented in Smart Buildings after this testing occurred – were not included in the completeness and expressiveness assessments, and was thus a mitigated but unavoidable limitation of this research. Further, the qualitative assessment relied on the small body of available beta documentation, publications, and sample implementations of ontology schemas. Finally, the fluid state of beta versions results in the potential for changes in ontology schema by contributors over the course of study, but this was preferred to using archived ontology versions, which would not lead to the most representative results as future implementations will not use such versions.

This research provides direction for the standardization of BAS ontology to support the development of Smart Building applications. These applications are ultimately valuable because they will simultaneously reduce building energy use and improve occupant comfort – this has been the goal of the Intelligent and Smart Building movements and continues to be the over-riding goal of automated control. As the requirements for Smart Building applications and ontology schemas change, these ontologies are in a fluid and evolving state, requiring constant reevaluation. As such an agreed upon standard would benefit from a modular approach that can integrate with other ontologies and a formalization of governance to clearly define and extend the ontology.  After reaching a consensus on the selection of an ontology standard to support Smart Building applications, the largest hurdle to mass adoption will be the implementation of ontologies in brownfield buildings. The open question of “how ontologies can be implemented for brownfield buildings with minimal manual effort while offering high quality data normalization?” persists in current discourse. This research has however provided a recommendation for the inclusion of Brick in the ontology standard to be used to support Smart Building applications.

**Acknowledgments**

This research was financially supported by the Natural Science and Engineering Research Council [CREATE 510284-2018], the Mitacs Accelerate program [IT15509], and Schneider Electric. The authors would specifically like to acknowledge the guidance and leadership of Oskar Nilsson and Jonas Bülow at Schneider Electric whose contributions were invaluable to this paper.